\documentclass[fleqn,usenatbib]{mnras}

\usepackage{newtxtext,newtxmath}
\usepackage[version=4]{mhchem}
\usepackage{comment}

\usepackage[T1]{fontenc}

\DeclareRobustCommand{\VAN}[3]{#2}
\let\VANthebibliography\thebibliography
\def\thebibliography{\DeclareRobustCommand{\VAN}[3]{##3}\VANthebibliography}

\usepackage{graphicx}	
\usepackage{amsmath}	
\usepackage{todonotes}

\title[Impact of C and CH reacting with H2]{Investigating the impact of reactions of C and CH with molecular hydrogen on a glycine gas-grain network}

\author[J.Heyl et al.]{
Johannes Heyl,$^{1}$\thanks{E-mail: johannes.heyl.19@ucl.ac.uk}
Thanja Lamberts,$^{2,3}$
Serena Viti$^{3,1}$
and Jonathan Holdship$^{3,1}$
\\
$^{1}$Department of Physics and Astronomy, University College London, Gower Street, WC1E 6BT, London, UK\\
$^{2}$Leiden Institute of Chemistry, Gorlaeus Laboratories, Leiden University, PO Box 9502, 2300 RA Leiden, The Netherlands \\
$^{3}$Leiden Observatory, Leiden University, PO Box 9513, 2300 RA Leiden, The Netherlands\\
}

\date{Accepted XXX. Received YYY; in original form ZZZ}

\pubyear{2015}

\begin{document}
\label{firstpage}
\pagerange{\pageref{firstpage}--\pageref{lastpage}}
\maketitle

\begin{abstract}
The impact of including the reactions of C and CH with molecular hydrogen in a gas-grain network is assessed via a sensitivity analysis. To this end, we vary 3 parameters, namely, the efficiency for the reaction \ce{C + H2 -> CH2}, and the cosmic ray ionisation rate, with the third parameter being the final density of the collapsing dark cloud. A grid of 12 models is run to investigate the effect of all parameters on the final molecular abundances of the chemical network. 
We find that including reactions with molecular hydrogen alters the hydrogen economy of the network; since some species are hydrogenated by molecular hydrogen, atomic hydrogen is freed up. 
The abundances of simple molecules produced from hydrogenation, such as \ce{CH4}, \ce{CH3OH} and \ce{NH3}, increase, and at the same time, more complex species such as glycine and its precursors see a significant decrease in their final abundances. We find that the precursors of glycine are being preferentially hydrogenated, and therefore glycine itself is produced less efficiently. 
\end{abstract}




\section{Introduction}
Interstellar dust plays a significant role in the rich chemistry that takes place in the interstellar medium. It is widely believed that complex-organic molecules (COMs) form on interstellar dust \citep{Herbst_Dishoeck, Caselli_review} since for certain molecules, grain-surface reactions are more efficient than gas-phase reactions. This is particularly important in cold astronomical environments where some gas-phase reactions may be highly inefficient, because a "third body" is needed to take up the excess heat of an exothermic reaction. Dust grains thus act as an energy sink allowing the chemistry to thrive and this can lead to the formation of more complex organic molecules. 

Both experimental work and modelling has shown that one such molecules, namely the amino acid glycine can be formed through energetic processing of the ices during the warm-up phase of star formation \citep{Bernstein, Woon, Lee,   Bossa, Cielsa,Garrod_glycine, Sato}, although there is evidence to suggest that glycine would undergo destruction under increased irradiation \citep{Pernet, Mate}. In addition, in a joint experimental and modeling effort, \cite{Ioppolo} suggested that non-energetic mechanisms such as atom-addition reactions might be a promising route for glycine formation. 

A new grain-surface reaction, inserting C atoms in \ce{H2} to form \ce{CH2} via \ce{C + H2 -> CH2}, was recently proposed to be barrierless by \cite{Simoncic}, based on earlier lab work by \cite{Krasnokutski_experimental_work}. They included this reaction in their network and found a far more rapid conversion of C to \ce{CH4}. Subsequently, \cite{Lamberts_methane} performed a combined experimental and computational work to investigate the importance of reactions with molecular hydrogen for the formation of methane. It was found that while the former reaction might not be fully barrierless, and the barrier likely depends on the binding site, the reaction \ce{CH + H2 -> CH3} does in fact proceed without a barrier. The reason these `dihydrogenation' reactions might be of interest is that they make \ce{H2} more chemically active, the importance of which was recognized already by \cite{Hasegawa} and by \cite{Meisner} in the context of water formation. Typically, \ce{H2} has one of the lowest binding energies of grain-surface species, lower than even atomic H \citep{H_binding_energy, Wakelam, H2_binding_energy}, which allows the molecule to diffuse readily on the surface. Moreover, the molecular hydrogen abundance in molecular clouds and pre-stellar cores is much higher than that of atomic hydrogen \citep{Dishoeck:1988,Goldsmith:2005}.

By including these reactions in chemical models, one might first of all expect changes in the \ce{CH4} abundance, but it is equally interesting to consider the effect on downstream species such as complex organic molecules, whose typical abundances are far lower. Their sensitivity to new reactions should be considered, as their more abundant precursors might see changes in their abundances. 

In this work, we look to build on the work by \cite{Simoncic} and \cite{Lamberts_methane} to investigate the impact of the dihydrogenation reactions of C and CH on our gas-grain chemical network. In particular, we are interested in observing the effect these reactions have on the production of glycine and its precursors. Our glycine network is based on the kinetic Monte Carlo network used in \cite{Ioppolo}, using in part updated rate constants from recent literature, as indicated in Table~\ref{ioppolo_reaction_network_table}.

We start by describing  the astrochemical model, our choice of parameters and how we evaluate the network sensitivity in Section \ref{methodology}. We then discuss the results as well as the astrochemical implications in Section \ref{results} and summarize our conclusions in \ref{conclusion}.
\section{Methodology}\label{methodology}
\begin{table*}
\label{ioppolo_reaction_network_table}
 \begin{tabular}{||c c c||} 
 \hline
 Reaction No. & Reaction  & Reference\\ [0.5ex] 
 \hline
 \hline
 1 & \ce{CO + OH -> HOCO} & \cite{Arasa2013} \\
 \hline
 2 & \ce{HOCO + H -> H_{2} + CO_{2}} & \cite{Goumans2008} \\
 \hline
 3 & \ce{HOCO + H -> HCOOH} & \cite{Goumans2008,Ioppolo2011} \\
 \hline
 4 & \ce{CH_{4} + OH -> CH_{3} + H_{2}O} & \cite{Lamberts2017b} \\
 \hline
 5 & \ce{NH_{2} + CH_{3} -> NH_{2}CH_{3}} & \cite{Ioppolo} \\
 \hline
 6 & \ce{NH_{3} + CH -> NH2CH_{2}} & \cite{Balucani2009} \\
 \hline
 7 & \ce{NH2CH_{2} + H -> NH_{2}CH_{3}} & \cite{Ioppolo} \\
 \hline
 8 & \ce{NH_{2}CH_{3} + H -> NH2CH_{2} + H_{2}} & \cite{Oba2014} \\
 \hline
 9 & \ce{NH_{2}CH_{3} + OH -> NH2CH_{2} + H_{2}O} & \cite{Ioppolo} \\
 \hline
 10 & \ce{NH2CH_{2} + HOCO -> NH_{2}CH_{2}COOH} & \cite{Woon} \\
 \hline
 11 & \ce{H_{2} + OH -> H_{2}O + H} & \cite{Meisner} \\
 \hline
 12 & \ce{O_{2} + H -> HO_{2}} & \cite{Lamberts2013} \\
 \hline
 13 & \ce{HO_{2} + H -> OH + OH} & \cite{Lamberts2013} \\
 \hline
 14 & \ce{HO_{2} + H -> H_{2} + O_{2}} & \cite{Lamberts2013} \\
 \hline
 15 & \ce{HO_{2} + H -> H_{2}O + O} & \cite{Lamberts2013} \\
 \hline
 16 & \ce{OH + OH -> H_{2}O_{2}} & \cite{Lamberts2013} \\
 \hline
 17 & \ce{OH + OH -> H_{2}O + O} & \cite{Lamberts2013} \\
 \hline
 18 & \ce{H_{2}O_{2} + H -> H_{2}O + OH} & \cite{Lamberts2017} \\
 \hline
 19 & \ce{N + O -> NO} & \cite{Ioppolo} \\
 \hline
 20 & \ce{NO + H -> HNO} & \cite{Fedoseev2012} \\
 \hline
 21 & \ce{HNO + H -> H_{2}NO} & \cite{Fedoseev2012} \\
 \hline
 22 & \ce{HNO + H -> NO + H_{2}} & \cite{Fedoseev2012} \\
 \hline
 23 & \ce{HNO + O -> NO + OH} & \cite{Ioppolo} \\
 \hline
 24 & \ce{HN + O -> HNO} & \cite{Ioppolo} \\
 \hline
 25 & \ce{N + NH -> N_{2}} & \cite{Ioppolo} \\
 \hline
 26 & \ce{NH + NH -> N_{2} + H_{2}} & \cite{Ioppolo} \\
 \hline
 27 & \ce{C + O -> CO} & \cite{Ioppolo} \\
 \hline
 28 & \ce{CH_{3} + OH -> CH_{3}OH} & \cite{Qasim2018} \\
 \hline
 \hline
 29 &\ce{C + H2 -> CH2} & \cite{Simoncic,Lamberts_methane}\\
 \hline
 30 & \ce{CH + H2 -> CH3} & \cite{Lamberts_methane}\\
\hline
\end{tabular}
\caption{Table of the reactions added to the standard UCLCHEM network.}
\end{table*}

\begin{table*}
\begin{tabular}{|l|l|l|}
\hline
\textbf{Parameter} &
  \textbf{Values} &
  \textbf{Comment} \\ \hline
Final Density of Phase 1 and Initial Density of Phase 2 &
  $10^{5}$ cm$^{-3}$, $10^{6}$ cm$^{-3}$, $10^{7}$ cm$^{-3}$ & Final density of Phase 1 same as initial density of Phase 2
   \\ \hline
\textbf{Efficiency} for barrierless $\ce{C + H_{2} -> CH_{2}}$ & 0, 0.05, 1 & \textbf{Efficiency} of 0 is equivalent to reaction being excluded. \\ \hline
Cosmic Ray Ionisation Rate &
  $\zeta, 10\zeta$ &
  $\zeta$ is the standard cosmic ray ionisation rate of $1.3 \times 10^{-17}$ s$^{-1}$ \\ \hline
\end{tabular}
\caption{The parameters that were varied in this work to assess the effect of the two reactions. }
\label{parameter_table}
\end{table*}

\subsection{The Astrochemical Model}
In this work, the gas-grain chemical code UCLCHEM was used \citep{UCLCHEM}\footnote{https://uclchem.github.io/}. UCLCHEM makes use of a rate equation approach to modelling the gas and grain-surface and bulk abundances.  The gas-phase reaction network is taken from the UMIST database \citep{UMIST}. The grain-surface network used was the default one as available on GitHub. 

Various reaction mechanisms are implemented in UCLCHEM. The grain-surface reaction mechanisms that exist in UCLCHEM include the Eley-Rideal mechanism as well as the Langmuir-Hinshelwood diffusion mechanism, which were implemented in \cite{Quenard}, as was the competition formula from \cite{Chang} and \cite{Garrod_CO2}. The binding energies that are used to calculate the diffusion reaction rate are taken from \cite{Wakelam}. We also included an updated version of the glycine grain-surface network from \cite{Ioppolo}, also including both the reactions \ce{C + H2 -> CH2} and \ce{CH + H2 -> CH3} as summarized in Table~\ref{ioppolo_reaction_network_table}. Note that the reaction \ce{OH + H2 -> H2O + H} had been already included, based on previous work by, e.g., \cite{Meisner}. The code also includes thermal and non-thermal desorption, such as due to \ce{H2} formation, cosmic ray ionisation as well as UV-induced desorption. Note that the astrochemical model used in \cite{Ioppolo} makes use of the non-diffusive grain-surface chemistry that is described in \cite{Garrod_CO2} and \cite{Jin_Garrod}. This is not used in UCLCHEM. The implications of this will be discussed later in this work.

UCLCHEM is used to model two distinct phases of the star formation process. Phase 1 is the free-fall collapse phase of a dark cloud for a default value of 5 million years, whereas Phase 2 models the warm-up phase immediately following Phase 1, with the initial density of Phase 2 equal to the final density of Phase 1. Phase 2 runs for 1 million years. Further details of the code can be found in \cite{UCLCHEM}. 

\subsection{Parameter Selection}
To assess the importance of the two proposed reactions to the network under various interstellar conditions, three parameters were varied, as listed in Table \ref{parameter_table}. The standard cosmic ray ionisation rate in UCLCHEM is $\zeta = 1.3 \times 10^{-17}$ s$^{-1}$. This is in line with typical values that are of the order $10^{-17}$ s$^{-1}$ in diffuse ISM conditions \citep{cosmic_ray1, cosmic_ray2, cosmic_ray3, cosmic_ray_review}. However, there exist observations of higher cosmic ray ionisation rates \citep{cosmic_ray5, cosmic_ray4}, which is why we also include analysis of a region with cosmic ray ionisation rate of 10$\zeta$. Cosmic ray ionisation is typically expected to break larger molecules into smaller radicals. We did not consider lower values of the cosmic ray ionisation rate, as these are typically not observed. The cosmic ray dependency on column density in \cite{Ross_paper} covered a range of values that were, however, already covered by the factor of 10 we consider here. While they found differences for lower densities during the collapse phase, these were ironed out once the collapse reached larger final densities, which is why here we do not include this dependency on column density.

Three different astronomical regions were modelled: 
\begin{enumerate}
    \item a dark cloud with a final density of $10^{5}$ cm$^{-3}$ 
    \item a low-mass protostar with a final density of $10^{6}$ cm$^{-3}$
    \item a high-mass protostar, with a final density of $10^{7}$ cm$^{-3}$
\end{enumerate}  
The heating profiles during Phase 2 for the last two cases are based on \cite{UCLCHEM_heating} and differ for each astronomical object. The dark cloud simulation was only run for Phase 1, but was allowed to run for a further  million years to allow the chemistry to settle. 

Another parameter that was varied was the efficiency, $\alpha$, of the extent to which the reaction \ce{C + H2 -> CH2} is barrierless. While \cite{Simoncic} considered the reaction to be fully barrierless, \cite{Lamberts_methane} found that the reaction barrier likely depends on the binding site. As such, our grid of models considers \textbf{efficiencies} for the reaction of 0 (the reaction is not included), 0.05 (5\% of binding sites lead to a barrierless reaction and 95\% of the binding sites have an infinitely high barrier) and 1 (the reaction is fully barrierless). What this means practically is that the reaction rate is multiplied by the efficiency. The reaction \ce{CH + H2 -> CH3} was included as only being barrierless, based on \cite{Lamberts_methane}.

\subsection{Evaluating the network sensitivity}
We quantify the effect of the new reactions on the model by considering the change in abundances of the species that are the most affected when taking the ratio of the abundances of the modified and original models. The modified model is the chemical network which has $\alpha = 1$, whereas the original model was taken to be the network which had neither of the dihydrogenation reactions. These two scenarios were taken to be the extremes of the parameter range in terms of including these reactions. The ratio is most sensitive to strong deviations in the molecular abundances as a result of the dihydrogenation reactions.  

This ratio is defined for each species $i$ as: 

\begin{equation}
\delta_{i}(t) = \frac{x_i^{M}(t)}{x_{i}^{O}(t)}, 
\end{equation}

where $x_{i}^{M}(t)$ is the abundance of species $i$ in the modified model at time $t$ and $x_{i}^{O}(t)$ is the abundance of the same species in the original model at time $t$.

We only considered species which had a value above a ``threshold of detectability". This was to ensure that we did not look at species whose original and changed abundances were below what can be observed from an astronomical point-of-view. For grain-surface species this threshold was set to $10^{-8}$ with respect to hydrogen whereas for gas-phase species this threshold equalled $10^{-12}$ with respect to hydrogen. We took $10^{-8}$ as a lower-limit threshold for grain-surface species, as this was the order of magnitude of the lowest reported abundances in \cite{Boogert}. Similarly, the gas-phase threshold was taken based on the abundances of COMs typically observed in the gas-phase, such as in \cite{l1544, l1498}.

We can also define a quantity that tracks the absolute change in the abundance of species: 

\begin{equation}
\Delta_{i}(t) = x_i^{M}(t) - x_{i}^{O}(t) = x_{i}^{O}(t)[\delta_{i}(t)-1], 
\end{equation}

This value indicates how species with relatively large abundances, such as elemental species or their hydrogenation products, are re-distributed.

\section{Results and Astrochemical Implications}\label{results}

\begin{table*}
\begin{tabular}{lllllllll}
\hline
\multicolumn{3}{|l|}{\textbf{Dark Cloud}}                 & \multicolumn{3}{l|}{\textbf{Low-Mass Star}} & \multicolumn{3}{l|}{\textbf{High-Mass Star}} \\ \hline
\multicolumn{1}{|l}{Species} & \multicolumn{1}{l|}{$\delta$} & \multicolumn{1}{l|}{Original Abundances} & Species     & \multicolumn{1}{l|}{$\delta$}  & \multicolumn{1}{l|}{Original Abundances} & Species     & \multicolumn{1}{l|}{$\delta$}  & \multicolumn{1}{l|}{Original Abundances}   \\ \hline
\#CH2 & 2.8  & 4.1 $\times 10^{-7}$  & \#CH2 & 2.8 & 4.1 $\times 10^{-7}$ & \#CH2  & 2.8  & 4.1 $\times 10^{-7}$ \\
\#CH3 & 2.3  & 2.6 $\times 10^{-7}$  & \#CH3 & 2.3 & 2.6$\times 10^{-7}$& \#CH3  & 2.3 & 2.6 $\times 10^{-7}$ \\
\#CH4 & 1.3 & 4.0 $\times 10^{-6}$& \#CH4 &  1.3 & 3.8 $\times 10^{-6}$ & \#CH4 & 1.3 & 3.8 $\times 10^{-6}$\\
\#NH3 & 1.1 & 3.8 $\times 10^{-6}$& \#NH3 &  1.1 & 3.7 $\times 10^{-6}$& \#NH3 & 1.1 & 3.7 $\times 10^{-6}$ \\
\#H2CS & 1.1 & 2.4 $\times 10^{-8}$& \#H2CS & 1.1  & 2.4 $\times 10^{-8}$& \#H2CS & 1.1 & 2.4 $\times 10^{-8}$\\
\#CH3OH & 1.04 & 1.5 $\times 10^{-5}$ & \#CH3OH &  1.04  & 1.3 $\times 10^{-5}$& \#CH3OH &  1.04  & 1.3 $\times 10^{-5}$\\
\#HNC & 1.03 & 2.3 $\times 10^{-8}$ & \#HNC &  1.04  & 2.3 $\times 10^{-8}$ & \#HNC &  1.04  & 2.3 $\times 10^{-8}$\\
\#H2SiO & 1.03 & 3.3 $\times  10^{-7}$ & \#H2SiO &  1.03  & 1.1 $\times 10^{-8}$ & \#H2SiO &  1.03  & 3.4 $\times 10^{-7}$\\
\#HCN & 1.02 & 1.7 $\times 10^{-7}$ & \#HO2 &  1.03  & 2.3 $\times 10^{-7}$& \#HO2 &  1.03  & 2.3 $\times 10^{-7}$\\
\#O2 & 1.02 & 1.8 $\times 10^{-6}$ & NO &  1.03  & 1.0 $\times 10^{-10}$& \#HCN & 1.02 & 1.6 $\times 10^{-7}$\\
\hline
\#CH & 1.1 $\times 10^{-15}$ & 7.2 $\times 10^{-7}$ & \#CH & 2.0 $\times 10^{-15}$  & 7.2 $\times 10^{-7}$ & \#CH  & 2.1  $\times 10^{-15}$ & 7.2 $\times 10^{-7}$\\
\#C & 2.4 $\times 10^{-13}$ & 1.4 $\times 10^{-6}$& \#C & 2.5 $\times 10^{-13}$ & 1.4 $\times 10^{-6}$& \#C & 2.5 $\times 10^{-13}$ & 1.4 $\times 10^{-6}$ \\
\#NCH4 & 3.7 $\times 10^{-13}$& 1.5 $\times 10^{-7} $& \#NCH4 & 3.4 $\times 10^{-13}$ & 1.5 $\times 10^{-7}$ & \#NCH4 & 3.4 $\times 10^{-13}$ & 1.5 $\times 10^{-7}$\\
\#NH2CH3 & 8.0 $\times 10^{-13}$ & 1.9 $\times 10^{-7} $& \#NH2CH3 & 8.3 $\times 10^{-13}$ & 2.0 $\times 10^{-7}$ & \#NH2CH3 & 8.3 $\times 10^{-13}$ & 2.0 $\times 10^{-7}$ \\
NH2CH3 & 1.5 $\times 10^{-12}$ & 8.7 $\times 10^{-10}$ & \#Si & 0.98 & 5.6 $\times 10^{-8}$ & \#Si & 0.98 & 5.6 $\times 10^{-8}$ \\
CH & 0.96 & 9.3 $\times 10^{-10}$& \#SiH & 0.99 & 2.5 $\times 10^{-8}$ & \#SiH & 0.99 & 2.5 $\times 10^{-8}$\\
CH3 & 0.98 & 1.5 $\times 10^{-9}$& \#SiH2 & 0.99 & 1.3 $\times 10^{-8}$& \#SiH2 & 0.99 & 1.3 $\times 10^{-8}$ \\
\#Si & 0.98 & 5.7 $\times 10^{-8}$ &  \#O & 0.99 & 7.8 $\times 10^{-5}$ & \#SI & 0.99 & 6.7 $\times 10^{-5}$ \\
\#SiH & 0.99 & 2.6 $\times 10^{-8}$&  \#H3CO & 0.99 & 1.7 $\times 10^{-6}$& \#H3CO & 0.99 & 1.7 $\times 10^{-6}$ \\
\#SiH2 & 0.99 & 1.4 $\times 10^{-8}$& \#HNO & 0.99 & 1.2 $\times 10^{-5}$&   \#HNO & 0.99 & 1.2 $\times 10^{-5}$\\ \hline
\end{tabular}
\caption{Summary of the species that experienced the greatest increases (top section) and decreases (bottom section) for each of the three astronomical objects in Phase 1. Species with a "\#" are grain-surface species. All other species are gas-phase.}
\label{phase1_final_changes}
\end{table*}

\begin{table*}
\begin{tabular}{llllll}
\hline
 \multicolumn{3}{l|}{\textbf{Low-Mass Star}} & \multicolumn{3}{l|}{\textbf{High-Mass Star}} \\ \hline
Species     & \multicolumn{1}{l|}{$\delta$}  & \multicolumn{1}{l|}{Original Abundances} & Species & \multicolumn{1}{l|}{$\delta$}  & \multicolumn{1}{l|}{Original Abundances}   \\ \hline
HOCO & 3.7 & 9.3 $\times 10^{-10}$ & HOCO & 2.1 & 4.3 $\times 10^{-8}$\\
\ce{H2O2} & 2.6 & 4.3 $\times 10^{-9}$ & CH3OH & 2.0 & 1.8 $\times 10^{-9}$ \\
\ce{CH3CHO} & 2.2 & 1.0$\times 10^{-7}$ & CH3CHO & 2.0 & 1.5 $\times 10^{-7}$ \\
\ce{CH3OH} & 2.1 & 3.7 $\times 10^{-9}$ & C2H4 & 2.0 & 2.5 $\times 10^{-9}$ \\
CH3CN & 1.7 & 1.0 $\times 10^{-9}$ & CH2CO & 1.9 & 1.8 $\times 10^{-10}$ \\
C4H & 1.6 & 3.2 $\times 10^{-10}$ & H2CO & 1.7 & 9.3 $\times 10^{-9}$\\
C3H2 & 1.5 & 5.6 $\times 10^{-9}$ & CH3 & 1.7 & 1.1$\times 10^{-10}$ \\
CH3CCH & 1.5 & 2.4$\times 10^{-8}$ & NH3 & 1.6 & 1.3 $\times 10^{-8}$\\
NH3 & 1.5 & 2.7 $\times 10^{-7}$ & CH3CN & 1.5 & 7.1 $\times 10^{-10}$\\
NH2CHO & 1.4 & 2.7 $\times 10^{-7}$ & C2H2 & 1.5 & 1.1 $\times 10^{-8}$  \\
\hline
NCH4 & 3.8 $\times 10^{-5}$ & 9.2 $\times 10^{-7}$ & NCH4 & 4.7 $\times 10^{-5}$ & $8.3 \times 10^{-7}$\\ 
NH2CH3 & 2.4 $\times 10^{-3}$ & 1.6 $\times 10^{-7}$ & NH2CH3 & 2.5 $\times 10^{-3}$ & 1.7 $\times 10^{-7}$\\
NH2CH2COOH & 6.0 $\times 10^{-2}$ & 6.3 $\times 10^{-9}$ & NH2CH2COOH & 6.3 $\times 10^{-3}$ & 7.2 $\times 10^{-8}$\\
H2S & 0.88 & 2.0 $\times 10^{-9}$ & NO & 0.82 & 4.0 $\times 10^{-6}$\\
SO2 & 0.92 & 4.4 $\times 10^{-8}$ & NCCN & 0.96 & $3.9 \times 10^{-7}$\\ 
MG+ & 0.93 & 8.0 $\times 10^{-8}$ & O2 & 0.96 & 7.1 $\times 10^{-6}$\\
O & 0.95 & $1.3 \times 10^{-5}$ & HCOO & 0.96 & $1.9 \times 10^{-10}$\\
CH2OH & 0.95 & 6.4 $\times 10^{-8}$ & C2N & 0.97 & 3.5 $\times 10^{-8}$ \\ 
O2 & 0.96 & 4.2 $\times 10^{-5}$ & O & 0.97 & $3.6 \times 10^{-8}$\\ 
SO & 0.97 & 1.9 $\times 10^{-6}$ & CO2 & 0.97 & $7.6 \times 10^{-6}$ \\ \hline
\hline
\end{tabular}
\caption{Summary of the species that experienced the greatest increases (top section) and decreases (bottom section) for each of the three astronomical objects in Phase 2. All species listed are gas-phase.}
\label{phase2_final_changes}
\end{table*}

We find that even though the amounts by which various species are affected differs for each stage of star formation, the general trends are broadly similar. As such, we group our analysis per phase. Tables \ref{phase1_final_changes} and \ref{phase2_final_changes} summarise the changes in terms of $\delta$. The effect of the enhanced cosmic ray ionisation rate is discussed in Section \ref{cosmic_ray_ionisation_rate}.

Our results differ from \cite{Ioppolo} in that, while glycine does form on the grains, it does not do so in Phase 1, as UCLCHEM does not utilise  non-diffusive grain-surface mechanisms. Instead, glycine forms on the grains as the temperature increases in Phase 2. 

\subsection{Impact of the Parameters}
In this sub-section we consider the role that the physical and chemical parameters play. Tables \ref{phase1_final_changes} and \ref{phase2_final_changes} show the changes in abundance when we compare the original network without the dihydrogenation reactions with the $\alpha=1$ case. Figures \ref{phase1_glycine_precursors} and \ref{phase2_glycine_precursors} show the time series of the abundances for glycine and its precursors.

\subsubsection{Final Density}

The final density of the collapsing cloud had a minor effect on the final abundances of the species in Phase 1. For all three astronomical objects modelled in Phase 1, we observe a significant decrease of grain-surface CH and C when the reactions are included and see an enhancement of grain-surface \ce{CH2}, \ce{CH3} and \ce{CH4}. However, the values of $\delta$ as well as their original abundances seem to be independent of the density, suggesting a saturation effect. 

In Phase 2, we observe that the final density of the collapsing cloud does affect the extent to which the added reactions influence the final abundances. We notice that several hydrogenation-based species have greater abundances at lower densities, including species such as \ce{HOCO}, \ce{H2O2}, \ce{CH3CCH} and \ce{H2CO}. 

\subsubsection{Efficiency}

For more abundant species, such as \ce{H2O} and \ce{CH3OH}, we find that the results obtained from using a branching fraction of 0.05 for the barrierless dihydrogenation of C are essentially the same as using a efficiency of 1 (the reaction is fully barrierless).  

We do find that the efficiency parameter plays a role in the final abundances of glycine and its precursors during the warm-up phase of low and high-mass stars. This can be seen in Figures \ref{phase1_glycine_precursors} and \ref{phase2_glycine_precursors}. For Phase 1, the species are not detectable except for the original configuration. However, we still observe that for the other three configurations  an increasing value of $\alpha$ corresponds to an increased level of depletion. In Phase 2, the configurations are all detectable and this same hierarchy remains in the gas-phase. 

\subsubsection{Cosmic Ray Ionisation Rate}\label{cosmic_ray_ionisation_rate}

The degree of cosmic ray ionisation is found to play an important role in enhancing or counteracting the role of the dihydrogenation reactions. The cosmic ray destruction routes we include in our standard network are from \cite{Garrod_CRP}. These consist of hydrogen abstraction reactions and reactions that produce radical-radical pairs of products. An enhanced cosmic ray ionisation rate leads to the destruction of many of  hydrogenated species, such as \ce{CH4}, \ce{NH3}, \ce{H2O} and \ce{CH3OH}, as well as their precursors. This leads to further hydrogen reservoirs being released and radicals being formed which can go on to form glycine and its precursors. Because no cosmic ray destruction mechanisms for these complex, larger, species are included, we find that these are more abundantly produced. 

This is important to consider in the context of glycine. In Figures \ref{phase1_glycine_precursors} and \ref{phase2_glycine_precursors}, we plot the time dependence of the abundance of glycine precursors for eight different parameter sets, including the enhanced cosmic ray ionisation rate. In Phase 1, we find that on the grains, the enhanced cosmic ray ionisation rate depletes the species. In Phase 2, the effect varies by configuration and species. The original configuration consistently leads to a decrease of all plotted species in the presence of enhanced cosmic ray ionisation. The $\alpha = 0$ configuration is depleted for the methylamine radical and glycine, but enhanced for methylamine. The $\alpha = 0.05$ and $\alpha = 1$ configurations are depleted for methylamine and glycine, but enhanced for the methylamine radical.

\subsection{General Implications}
As can be seen in Tables \ref{phase1_final_changes} and \ref{phase2_final_changes}, the inclusion of reactions with molecular hydrogen affects the hydrogen economy of the reaction network. Previously, the reaction network had a significant amount of \ce{H2} being adsorbed or produced on the surface with no chemical destruction mechanisms. The \ce{H_{2}} molecules are a previously untapped hydrogen reservoir that is now being utilised \citep{Hasegawa}. Because one \ce{H2} frees up two H atoms on the surface, other atomic hydrogenation reactions can take place more easily. Therefore, we observe the increase in the abundances of species in Phases 1 and 2 that are the products of hydrogenation. While for many of the more common species, the relative increase, i.e., $\delta$ is small, the abundance increases in absolute terms. There are large relative and absolute changes in the network of less abundant species, such as \ce{NH2CH2, NH2CH3 and NH2CH2COOH} and there are fairly large absolute changes in the network of highly abundant species, such as C and its hydrogenation products.

We can also comment on the carbon budget. The previously defined $\Delta$ parameter allows us to consider how carbon is redistributed as a result of the new reactions being included. For instance, for the dark cloud during Phase 1, the total $\Delta$ for the main carbon-based grain-surface species that increase $$\Delta_\text{total}(\#\ce{CH2} +  \#\ce{CH3} + \#\ce{CH4} + \#\ce{H2CS} + \#\ce{CH3OH}) = 2.9 \times 10^{-6}\;.$$ is nearly equal to that of the total decrease $\Delta$ of main grain-surface species: $$\Delta_\text{total}(\#\ce{C} + \#\ce{CH} + \#\ce{NCH4} + \#\ce{NH2CH3} = 2.5 \times 10^{-6}\;.$$ From this we can see that the dihydrogenation reactions redistribute the carbon between the aforementioned species. The remaining carbon is redistributed to other species in the network in smaller amounts. We also observe that besides the methyl radical, also species that contain the \ce{CH3} group, such as \ce{CH3OH} and \ce{CH3CN} see increases in their abundances, via the reactions \ce{CH3 + OH -> CH3OH} and \ce{CH3 + CN -> CH3CN}. 

In a similar fashion, nitrogen is redistributed throughout the network. The grain-surface ammonia abundance increases by 10\%, i.e., 3.8$\times 10^{-7}$. The decrease in \#\ce{NCH4} and \#\ce{NH2CH3} accounts for 3.4$\times 10^{-7}$ or $\sim 90\%$. 

\subsection{Implications for Simple Grain-Surface Species}
\begin{figure*}
\includegraphics[width=2.1\columnwidth]{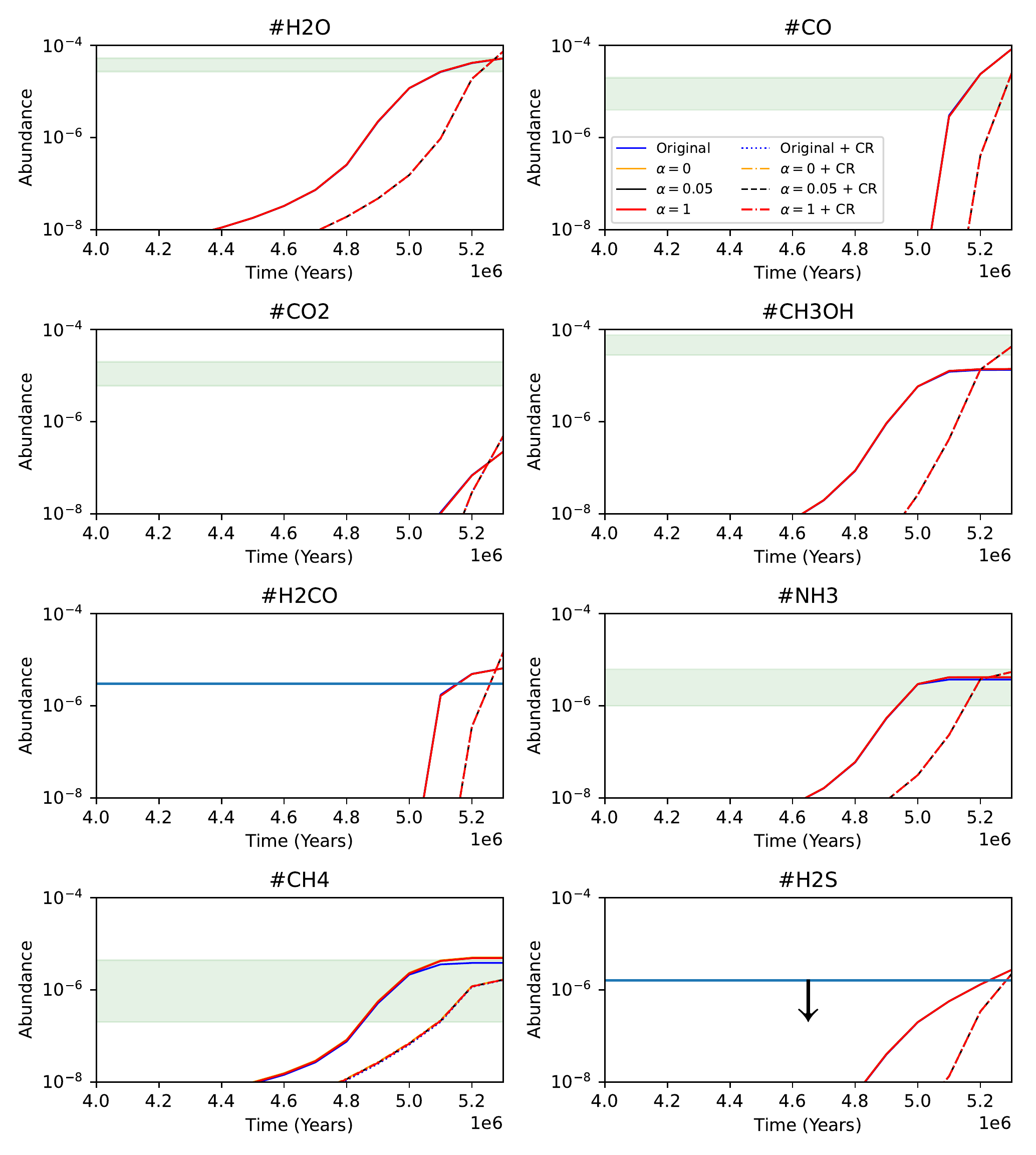}
\caption{Time series of the abundances of grain-surface \ce{H2O}, \ce{CO}, \ce{CO2}, \ce{CH3OH}, \ce{H2CO}, \ce{NH3}, \ce{CH4} and \ce{H2S} in Phase 1 of a dark cloud. We include the species that have securely identified or likely identified. The abundances were adapted from \protect\cite{Boogert}. The shaded areas include the $1\sigma$ region of abundances. In the case of \ce{H2CO}, no uncertainty was provided in the original source, so there is no shaded area. Grain-surface \ce{H2S} only has an upper limit on its abundance. For both normal and enhanced cosmic ray ionisation rates, the time-series differ very little, which is why it is difficult to distinguish them visually.}
\label{ice_species}
\end{figure*}

In the light of the recent ice observations with the James Webb Space Telescope, both published \citep{Yang:2022} and upcoming \citep{IceAge_proposal_1309}, it is important to consider the effect on the main ice constituents. Figure \ref{ice_species} shows the time-evolution of the abundances of grain-surface  \ce{H2O}, \ce{CO}, \ce{CO2}, \ce{CH3OH}, \ce{H2CO}, \ce{NH3} and \ce{CH4} in Phase 1 of a dark cloud. These are species that have been securely or likely identified in the ices \citep{Boogert}. The shaded areas in the plots indicate the 68\% confidence interval for the measured abundances, taken from \cite{Boogert}. In \cite{Boogert}, the abundances were given in terms of the median value as well as the upper and lower quartiles. It was assumed that the spread in the measurements was Gaussian, which meant that the interquartile range represented 1.36$\sigma$. This spread in measurements is due to both observational error and source-to-source variation. We observe that we recover the measured abundances for most of the species within the uncertainty, with the exception of grain-surface \ce{CO2}. The inclusion of the dihydrogenation reactions does not change how well the models agree with the abundance measurements, however, for all hydrogenation products we observe that the inclusion of reactions with molecular hydrogen increases their abundance, as a result of the additional atomic hydrogen on the surface. In short, despite uncertainties surrounding activation energies, networks and binding energies, we are able to recover observational abundances reasonably well when we include the reactions with molecular hydrogen and this gives us confidence that the predictions we make for glycine and its precursors are accurate. 

\subsection{Implications for Glycine and its Precursors}

In Tables \ref{phase1_final_changes} and \ref{phase2_final_changes}, we observe that the abundances of glycine and its precursors decreases if molecular hydrogen is part of the reaction network. We can also explain why the abundance of precursors of glycine, gas and grain \ce{NH2CH3} and \ce{NH2CH2} decrease. The former is formed through the reaction \ce{NH2 + CH3}, but since more atomic H is present on the grains, both radical species are preferentially hydrogenated. The inclusion of \ce{H2} as a reacting species, not just in the context of the two reactions we consider in this work, introduces greater competition for radicals that are needed for the formation of complex organic molecules. This results in the lower abundances of \ce{NH2CH3} and \ce{NH2CH2}.
\begin{figure*}
\includegraphics[width=2.1\columnwidth]{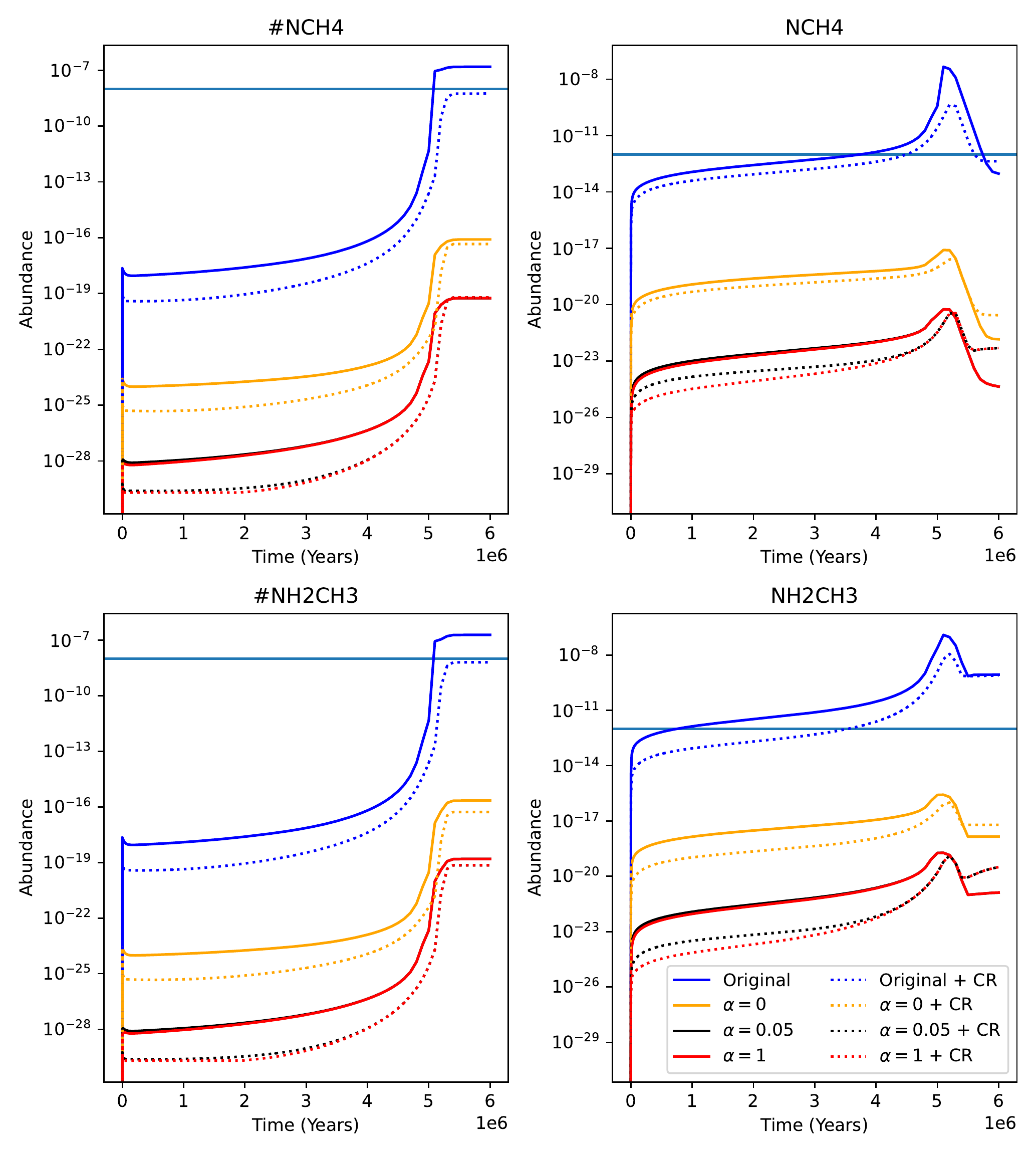}
\caption{Time series of the abundances of grain-surface and gas-phase \ce{NH2CH2} and \ce{NH2CH3} in Phase 1 of a dark cloud. Furthermore, we observe that the inclusion of the dihydrogenation reactions, regardless of efficiency $\alpha$ severely depletes the abundances of the glycine precursors in both phases relative to the original model which did not include either of the dihydrogenation reactions. Also plotted are the limits of detectability we have used for gas and grain-surface species. We do not plot glycine, as it is not formed at all in Phase 1. We observe that only the original model is capable of producing 'detectable' levels of methylamine and the methylamine radical. For the other configurations, an increase in $\alpha$ results in increased depletion of the species relative to the original model. We also observe that enhanced cosmic ray ionisation depletes the abundances on the grains but not in the gas.}
\label{phase1_glycine_precursors}
\end{figure*}

\begin{figure*}
\includegraphics[width=2.1\columnwidth]{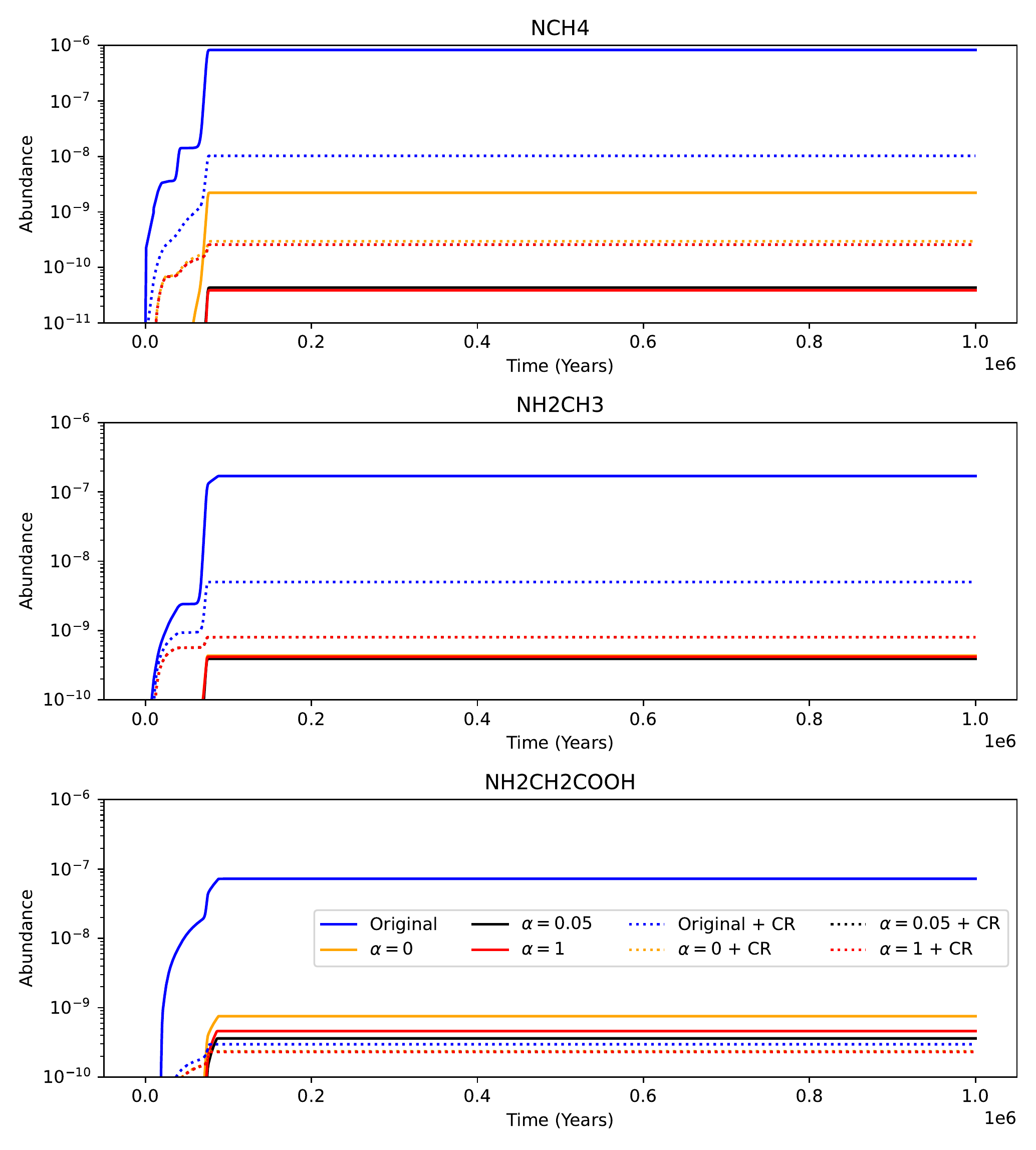}
\caption{Time series of the abundances of gas-phase \ce{NH2CH2}, \ce{NH2CH3} and \ce{NH2CH2COOH} in Phase 2 of a high-mass star. We observe that glycine is produced in the warm-up phase. The enhanced cosmic ray ionisation rate is found to significantly deplete all three species in the gas-phase for the original model. For \ce{NH2CH2} and \ce{NH2CH3}, when $\alpha = 0$, $\alpha=0.05$ or $\alpha=1$, the enhanced cosmic ray ionisation rate results in an increase of their abundances. For glycine, the enhanced cosmic ray ionisation rate seems to decrease its gas-phase abundance.}
\label{phase2_glycine_precursors}
\end{figure*}

We can also use this to justify the impact of the efficiency. Figures \ref{phase1_glycine_precursors} and \ref{phase2_glycine_precursors} plot the time series for the various efficiencies as well as with enhanced cosmic ray ionisation in Phase 1 and 2, respectively. We previously remarked that the original configuration produced the most of glycine and its precursors. For the other configurations, the greater the value of $\alpha$, the greater the depletion of these species. This makes sense when one considers that an increasing value of $\alpha$ results in more \ce{H2} being consumed and therefore more atomic H becoming available to hydrogenate precursors. 

We now look to compare our results with observations. We do this separately for glycine and its precursors. We also discuss the implications of not using  non-diffusive grain-surface mechanisms in our code, such as the ones discussed in \cite{Garrod_CO2} and \cite{Jin_Garrod}.

\subsubsection{Methylamine and the methylamine radical}
Methylamine (NH$_2$CH$_3$) and the methylamine radical  (NH$_2$CH$_2$) are important precursors of glycine. The hydrogen abstraction of methylamine to form the methylamine radical is crucial, as there is growing evidence to suggest that the reaction NH$_2$CH$_2$ + HOCO --> NH$_2$CH$_2$COOH is a feasible glycine formation route \citep{methylamine_radical_paper_nature}. Confirmed detections of methylamine in high-mass star forming regions are summarised in Table \ref{methylamine_abundance_table}. We observe improved level of agreement between our model outputs and observations when the reactions are included with $\alpha = 1$. We observed significant enhancement when the cosmic ray ionisation rate was increased. This suggests that if dihydrogen is chemically active on the grains, one would need to consider regions of high cosmic ray ionisation rate to detect these precursors of glycine, as these reactions reduce the abundance of methylamine. In the case of the \cite{methylamine_sagitarius} observation, we have confidence in the value of our ratio, as the chemical network for methanol is well-established.

However, the entirety of the above discussion regarding the agreement of our results with observations is incomplete without discussing the effect of the nondiffusive reaction mechanisms being absent in our modelling. These mechanisms are of particular use when considering reactions between reactants which are likely to react very slowly via the Langmuir-Hinshelwood diffusion mechanism, such as the reaction between CO and OH to form \ce{CO2}. Methylamine and the methylamine radical are formed via reactions 6 and 7, which involve species with high binding energies, thereby making their formation at 10K inefficient via diffusion. As a result, the fact that we do not include the non-diffusive mechanisms means that methylamine and its radical are under-produced.

\begin{table*}
 \begin{tabular}{||c c c c c||} 
 \hline
 Reference Molecule & Reference & Abundance Measurements (Relative to Reference Molecule)& Original Model Ratio & New Model Ratio\\ [0.5ex] 
 \hline
 \hline
 CH$_3$OH & \cite{methylamine_sagitarius} & $8 \times 10^{-3} - 0.1$ & 37 & 0.02 \\
 H$_2$ & \protect\cite{methylamine_abundance} & $1.5 \pm 1.1 \times 10^{-8}$ & $3.5 \times 10^{-7}$ & $3.9 \times 10^{-10}$ \\
 \hline
\end{tabular}
\caption{Table of methylamine abundance measurements relative to reference molecules for high-mass stars. Also included are the corresponding ratios obtained in this work for high-mass stars with the standard cosmic ray ionisation rate.}
\label{methylamine_abundance_table}
\end{table*}

\subsubsection{Glycine}
While there may be no confirmed detection for glycine in the literature, various estimates exist. In \cite{glycine_detection}, an upper limit of 0.3\% with respect to water was determined, whereas in \cite{Serena_glycine_paper}, this was estimated to be around 0.1\%. In this work, we find that when the dihydrogenation reactions are not included  this value is 0.07\% and when we include both reactions then it is   $2\times10^{-4} \%$. We should note that in the absence of experimentally-motivated gas-phase glycine destruction reactions the values derived in this work are only upper limits, if one neglects non-diffusive mechanisms. In the previous sub-section, we discussed that methylamine and its radical are underproduced. This will result in glycine being underproduced as well, not just due to the underproduction of its precursors, but also because reaction 10 is less efficient if assumed to be diffusion-only.

\section{Conclusion}\label{conclusion}
In this work, we considered the effect of including the reactions of \ce{H2} with \ce{C} and \ce{CH} in our grain-surface network. We ran a grid of 12 models that vary the final density of the collapsing cloud, the \textbf{efficiency} for the `barrier' of \ce{C + H2 -> CH2} as well as the cosmic ray ionisation rate. 

Making molecular hydrogen chemically active unlocks a previously untapped reservoir of hydrogen, and therefore freeing up the use of atomic hydrogen for hydrogenation reactions. A particularly interesting consequence of this is that making \ce{H2} more chemically active decreased the abundances of glycine and its precursors. This may aid in explaining why methylamine, the methylamine radical as well as glycine have remained undetected so far. 

We note that we  do not have a comprehensive gas-phase network for glycine and its precursors. That is likely to be a limitation. While it is still likely that glycine and its precursors form on the grains and then evaporate into the gas-phase, it is possible that there would be gas-phase destruction routes as well. Additionally, cosmic-ray ionisation destruction routes on the grains and in the gas-phase are likely also needed, as these typically break large molecules down into smaller radicals which are then recycled for further gas-phase reactions. As such, the abundances we obtain for glycine and its precursors are likely to only be upper limits. 

An additional limitation is the absence of the non-diffusive reaction mechanisms discussed in \cite{Garrod_CO2} and \cite{Jin_Garrod}. The consequence is that glycine and its precursors do not form efficiently on the grains at 10 K, which is different to what was found in \cite{Ioppolo}. As such, they are under-produced in our models, whereas diffusion-efficient reactions overproduce certain species. However, without implementing this formalism in the code, it is difficult to assess the relative impacts of these mechanisms on the final abundances.

\section*{Acknowledgements}
We thank the anonymous referee for their constructive comments that improved the quality of the manuscript. J. Heyl is funded by an STFC studentship in Data-Intensive Science (grant number ST/P006736/1). T. Lamberts is grateful for support from NWO via a VENI fellowship (722.017.008). This work was also supported by European Research Council (ERC) Advanced Grant MOPPEX 833460. S. Viti acknowledges support from the European Union’s Horizon 2020 research and innovation programme under the Marie Skłodowska-Curie grant agreement No 811312 for the project ``Astro-Chemical Origins” (ACO). 

\section*{Data Availability}

The data underlying this article are available in the article and in its online supplementary material.




\bibliographystyle{mnras}
\bibliography{example} 





\bsp	
\label{lastpage}
\end{document}